\documentclass[fleqn,twoside,twocolumn,nofootinbib,showkeys]{revtex4} 
\usepackage[nocpr]{ujp} 

\begin{document}
\title[Research Into a Neon Spectral Line Profile of Dusty Plasma]
{RESEARCH INTO A NEON\\ SPECTRAL LINE PROFILE OF DUSTY PLASMA}%
\author{A.A. Pikalev}
\affiliation{Petrozavodsk State University, Faculty of Physical Engineering}
\address{33, Lenin Str., 185910, Petrozavodsk, Republic of Karelia, Russia}
\email{pikalev@dims.karelia.ru}
\author{L.A.~Luizova}
\affiliation{Petrozavodsk State University, Faculty of Physical Engineering}
\address{33, Lenin Str., 185910, Petrozavodsk, Republic of Karelia, Russia}
\email{pikalev@dims.karelia.ru}

\udk{533.9.082.5} \pacs{52.27.Lw,  52.70.Kz} \razd{\secv}

\autorcol{A.A.\hspace*{0.7mm}Pikalev, L.A.\hspace*{0.7mm}Luizova}

\setcounter{page}{375}%

\begin{abstract}
Ordered dusty structures influence plasma conditions.\,\,This
influence can be revealed, when plasma spectral characteristics
change, as dusty particles are injected.\,\,For example, a variation
in the atomic temperature leads to a variation in the profiles of
spectral lines.\,\,We studied the profile of a 585~nm neon spectral
line in the dusty structures.\,\,The structures levitated in a
positive column of a glow discharge at a pressure of 50--150~Pa and
with a current of 1--9~mA.\,\,We scanned the profile with the use of
a Fabry--Perot interferometer, by changing the air pressure between
the interferometer mirrors.\,\,To process the data, a special
algorithm was developed.\,\,The algorithm is resistant to a noise
and a scanning speed instability.\,\,We have found an upper bound of
the impact of dusty structures on the profile width.\,\,The
appearance of macroparticles changes the atomic plasma temperature
less than by 10~K.
\end{abstract}

\keywords{spectral line profile, dusty plasma, dusty structure,
Fabry--Perot interferometer.}

\maketitle
\section{Introduction}
Charged macroparticles levitating in plasma can form ordered dusty
structures.\,\,Properties of the structu\-res are determined by
plasma conditions and {\it vice versa}.\,\,Electrons and ions
recombinate on the surface of macroparticles.\,\,Having an
electrical charge, the grains disturb an electrical field around
themselves.\,\,Changing the plasma spectral characteristics, as
dusty particles are injected, is well studied (see, e.g.,
\cite{Bulba}).

The impact of dusty structures on the profiles of spectral lines is
not studied enough.\,\,In \cite{Pal}, the spectral line profile
$H_\beta$ was modeled, and it was shown that an electric field
around dusty particles leads to the Stark \mbox{broadening.}

Elementary processes can also be reflected in spectral line
profiles.\,\,The spectral line profile $H_\alpha$ in an
$\mathrm{Ar}$--$\mathrm{C}_2\mathrm{H}_2$ RF discharge was
investigated in \cite{Stefanovic}.\,\,In plasma without particles,
there are hot H atoms with energies more than 10~eV.\,\,They are
reflected in the wings of a spectral line profile H$_\alpha$.\,\,The
hot atoms disappear, when dusty grains appear.

Gas heating is one more way for the dusty particles to change the
profiles of spectral lines.\,\,Changing the atomic temperature leads
to changing the profiles of spectral lines due to the Doppler
broadening.\,\,Electrons and ions recombine on the surfaces of
particles.\,\,Hence, an electric field increases to compensate
charge losses \cite{Vasilyak}.\,\,The surfaces of particles are by
tens Kelvins hotter than the gas around \cite{surface temperature}
and heat up the gas in the center of a tube where the light emission
is the most intensive.

We tried to detect the influence of dusty structures on the 585-nm neon
spectral  line in a glow discharge positive column.

\section{Setup}
To conduct the experiment, we constructed a setup consisting of a
complex  plasma-creating system \cite{Bulba}, a Fabry--Perot
interferometer, a monochromator, and a photoregistration system
based on a pho\-to\-mul\-tip\-lier.\,\,Light from a discharge tube
passed through the interferometer.\,\,An objective (focus length
\mbox{$F=10$}~cm) formed rings on a monochromator input slit.\,\,A
0.15-mm aperture detached a central part of the rings.\,\,Scanning
was performed by changing the pressure in a camera with the
interferometer.\,\,The experimental data were obtained with a
special computer \mbox{program.}

\begin{figure*}
\vskip1mm
\includegraphics[width=11.5cm]{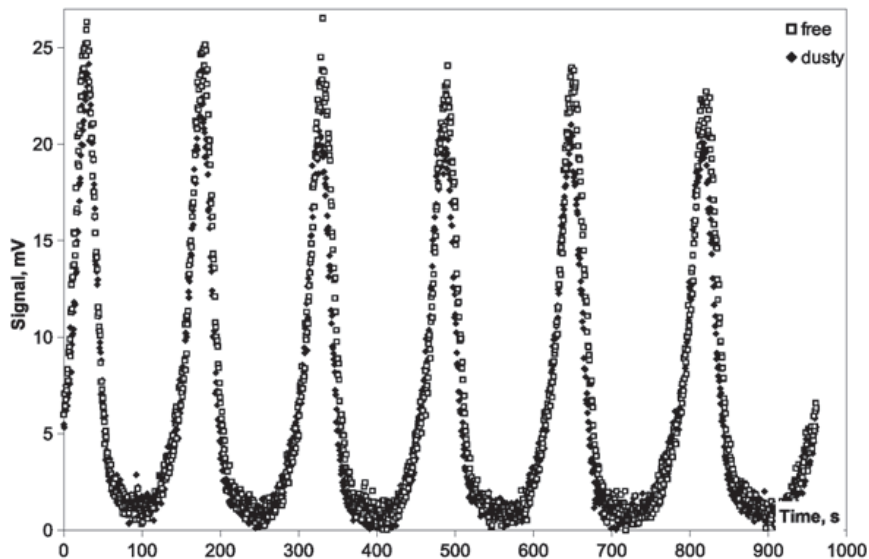}
\vskip-3mm \parbox{11.5cm}{\caption{\label{original} Raw
interferograms. $p=50$~Pa, $I=3$~mA}}
\end{figure*}

A DC glow discharge was in the tube 3~cm in dia\-me\-ter.\,\,After
injections of (1--10)-$\mu $m Zn particles, a dusty structure
levitated over a special narrowing of the positive column of the
discharge.\,\,The typical structure had a diameter of 3--5~mm and a
length of 7--10~mm.\,\,The width of the interferometer was 14~mm.
The objective and a lens between the tube and the interferometer
drew the discharge image on the slit.\,\,The central part of the
rings coincided with the dusty structure image.\,\,The monochromator
detached the 585.2-nm neon spectral line.

A signal from the photomultiplier was measured with a rate of $2
\times 10^{4}$ samples per second.\,\,Every experimental point was
obtained by averaging the samples in 600~ms.\,\,The typical
interferogram registration \mbox{took 1000~s.}

To compensate the floating of discharge conditions, we were changing
the registration order.\,\,If we registered the profiles of the free
plasma before the profiles of the plasma with particles for one day,
we worked for the next day under the same conditions, but in the
reverse order.

\section{Data Processing Algorithm}
The discharge current must be low to let the self-organization of
dusty structures.\,\,Hence, the light intensity is weak, and a
registration system noise is significant.\,\,One more problem is a
scanning speed instability.\,\,To compare the spectral line
profiles, we need to transform them.\,\,A special algorithm was
developed to solve the problems.

First of all, a dark signal was measured before and after every
interferogram  registration and was subtracted, considering a linear
change of the dark signal.\,\,The next step was to divide the
interferogram into individual profiles.\,\,Then we had to find the
maxima of profiles.\,\,A simple maximum of the signal could not be
used because of the noise.\,\,We approximated 15\% of the profile
top with a parabola and used its vertex as the profile maximum.

\begin{figure*}
\vskip1mm
\includegraphics[width=11.5cm]{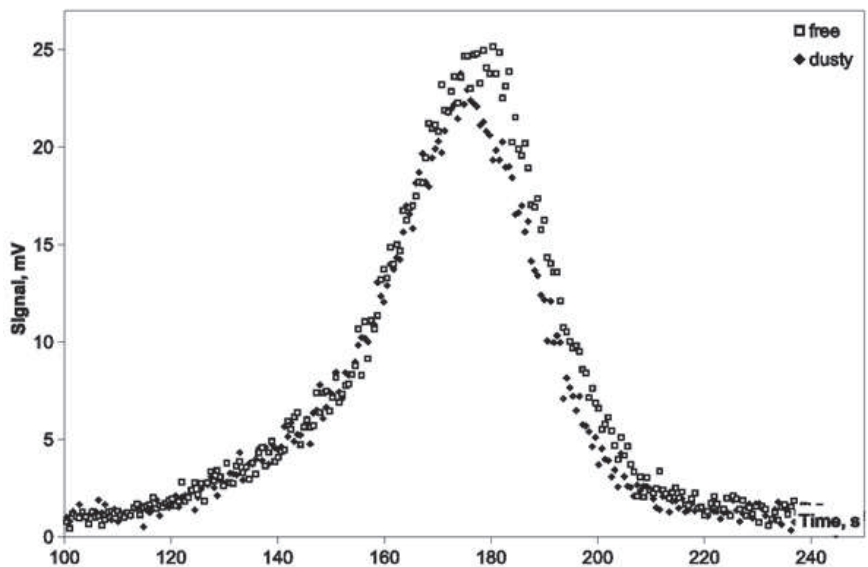}
\vskip-1mm \parbox{11.5cm}{\caption{\label{one_peak} One peak of the
interferograms presented in Fig.~\ref{original}}}
\end{figure*}

\begin{figure*}[!]
\vskip5mm
\includegraphics[width=11.5cm]{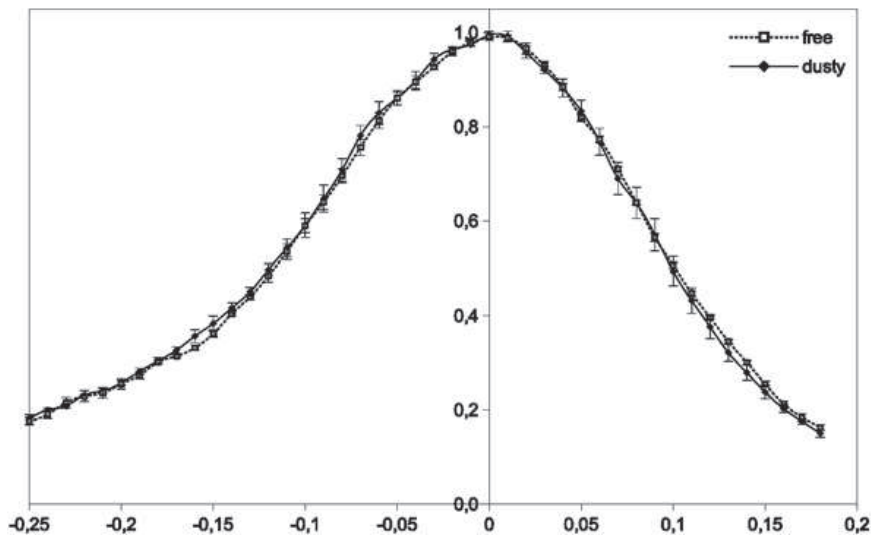}
\vskip-1mm \parbox{11.5cm}{\caption{\label{averaged} Averaged
profiles obtained from the interferograms shown in
Fig.~\ref{original}. Abscissas are in fractions of the
interferometer constant}}\vspace*{-0.5mm}
\end{figure*}

After that, we transformed abscissas.\,\,The interferometer constant
$\Delta \lambda = \frac{\lambda_0^2}{2d} = 12.2 $~pm ($\lambda_0 =$
$= 585.2$~nm~-- a center of the spectral line, $d=14 $~cm~-- a
distance between the interferometer mirrors) is scanned in time
between 2 neighbor interferogram maxima.\,\,The scanning speed was
not constant.\,\,We used fractions of $\Delta \lambda$ as new
coordinates $x$:
\[
x = c_0 + c_1 t + c_2 t^2.
\]
The coefficients $c$ were calculated for a current profile to have
$x=0$ for its  maximum, $x=-1$ for its left neighbor maximum and
$x=1$ for its right neighbor maximum.\,\,The profile must have
neighbors, that is why we could not transform coordinates of the
first and last profiles in each interferogram.\,\,Then we normalized
the ordinates of the profiles to their maxima taking a slight
intensity change into account.

Then we tried 2 ways of calculations.

-- To determine the profile width, we approximated points with
ordinates
 from 0.43 to 0.57 with straight lines and used abscissas of the points of lines
  with ordinates of 0.5.

-- To compare the shapes of profiles, we calculated the averaged profile for
 every interferogram.\,\,Every averaged profile consisted of points with a step of $0.01 \Delta
 \lambda$.\vspace*{-2mm}

\section{Results}
An example of raw interferograms is shown in
Figs.~\ref{original}--\ref{one_peak}.\,\,These interferograms after
transformation are presented in Fig.\,\,\ref{averaged}.\,\,The
difference between the original interferograms is caused by
variations of the scanning speed and the start moment.\,\,No changes
in the shapes of profiles caused by particles can be
detected.\,\,The profiles are asymmetric due to an isotopic
\mbox{structure.}

To detect changes in the widths of profiles, we calculated
differences of the widths of profiles with and without dusty
particles.\,\,After that, we checked a hypothesis that the
difference is equal to zero.\,\,The results are presented in
Table.\,\,It can be seen that zero is in the confidence interval for
every investigated \mbox{condition.}

\begin{table}[t]
\vskip4mm \noindent\caption{Difference between the widths\\ of
profiles with and without particles} \vskip4mm\tabcolsep4.7pt
\label{width_dif} \noindent{\footnotesize\begin{tabular}{|c|c|c|c|}
       \hline \multicolumn{1}{|c|}
{\rule{0pt}{5mm}Conditions} &\parbox[c][15mm][c]{11mm}{Number of the
profiles pairs} &\parbox[c][19mm][c]{17mm}{Average difference of the
widths, pm} &
\parbox[c][15mm][c]{22mm}{Confidence interval for~90\% probability,\\ pm}
\\[-0.5mm]
 \hline
        \rule{0pt}{5mm}~\,50~Pa, 1~mA & 19 & --0.019~\, & 0.032 \\
        ~\,50~Pa, 3~mA & 16 & --0.012~\, & 0.038 \\
        ~\,50~Pa, 6~mA & 21 &  0.003 & 0.019 \\
        ~\,50~Pa, 9~mA & 18 &  0.013 & 0.018 \\
        ~\,60~Pa, 3~mA & 10 & --0.031~\, & 0.044 \\
        100~Pa, 1~mA & 18 & --0.012~\, & 0.033 \\
        100~Pa, 3~mA & 37 & 0.007 & 0.031 \\
        100~Pa, 6~mA & 18 & --0.010~\, & 0.026 \\
        150~Pa, 3~mA & 19 & --0.008~\, & 0.042 \\[2mm]
        \hline
    \end{tabular}}\vspace*{2mm}
\end{table}

The confidence interval value is the upper bound of the dusty
influence on the profile.\,\,A simulation with a program
\cite{program} showed that the gas temperature change by 10 K leads
to the profile width change by 0.029~pm.\,\,For the simulation, we
used the Doppler profile shape and an ideal Fabry--Perot apparatus
function for the mirrors reflectivity $R = 0.8$.

If no change in profiles was detected, one can use gas temperature
data obtained for a discharge without particles for dusty plasma
studies.\,\,The presented algorithm will be useful for processing
the interferograms in a further research.

\vskip5mm
 \textit{We appreciate A.I.\,\,Scherbina for his assistance with the
experimental setup.}

\textit{The work was supported by the Ministry of Education and
Science of Russia,  grant No.\,\,14.B37.21.0755 and  the Program of
strategical development of PetrSU.}

\vspace*{-5mm} \rezume{А.А.\,Пiкалєв, Л.А.\,Луізова }{ДОСЛІДЖЕННЯ
КОНТУРА СПЕКТРАЛЬНОЇ\\ ЛІНІЇ НЕОНА У ПИЛОВІЙ ПЛАЗМІ} {Впорядковані
плазмово-пилові структури впливають на властивості плазми.\,\,Цей
вплив проявляється у зміні спектральних характеристик плазми при
внесенні частинок.\,\,Наприклад, зміна атомної температури приводить
до зміни контурів спектральних ліній.\,\,Ми досліджували контур
спектральної лінії неону 585,2~нм.\,\,Пилова структура зависала в
позитивному стовпі тліючого розряду при тиску 50--150~Па і струмі
1--9~мА.\,\,Сканування контуру проводилося за допомогою
інтерферометра Фабрі--Перо шляхом зміни тиску повітря між
дзеркалами.\,\,Для обробки даних було розроблено алгоритм, стійкий
до шумів і непостійності швидкості сканування.\,\,Було знайдено
верхню межу впливу пилових структур на ширину контуру спектральної
лінії в досліджених умовах.\,\,Поява пилу змінює температуру газу
менше ніж на 10~К.}

\end{document}